\documentclass[12pt]{iopart}
\usepackage{iopams}
\newcommand{\be}{\begin{equation}}
\newcommand{\ee}{\end{equation}}
\newcommand{\bee}{\begin{eqnarray}}
\newcommand{\eee}{\end{eqnarray}}
\newcommand{\x}{{\vec{\mbox{\boldmath{$x$}}}}}
\newcommand{\y}{{\vec{\mbox{\boldmath{$y$}}}}}

\newcommand{\vy}{{\dot\y}}
\newcommand{\p}{{\vec{\mbox{\boldmath{$p$}}}}}
\newcommand{\q}{{\vec{\mbox{\boldmath{$q$}}}}}

\newcommand{\F}{{\vec{\mbox{\boldmath{$F$}}}}}
\newcommand{\K}{{\mathcal{K}}}

\newcommand{\scl}{S_{\mathrm{cl.}}}

\newcommand{\dis}{\displaystyle}

\begin{document}

\jl{31}

\title[Particle Propagator and Path Integral Derivation]
	  {Particle Propagator in Elementary Quantum Mechanics:
	   a New Path Integral Derivation}

\author{Stefano Ansoldi\dag\footnote[3]{E-mail address: 
ANSOLDI@TRIESTE.INFN.IT},
        Antonio Aurilia\ddag\footnote[4]{E-mail address: 
AAURILIA@CSUPOMONA.EDU}
        and
        Euro Spallucci\dag\footnote[5]{E-mail address: 
SPALLUCCI@TRIESTE.INFN.IT}
       }
\address{\dag\
         Dipartimento di Fisica Teorica,
         Universit\`a di Trieste,\\
         Strada Costiera 11, 34014 Trieste,\\
         INFN, Sezione di Trieste
        }
\address{\ddag\
         Department of Physics,
         California State Polytechnic University,
         Pomona, CA 91768
        }

\begin{abstract}
This paper suggests a new way to compute the path
integral for simple quantum mechanical systems. The new algorithm
originated from previous research in string theory. However, its essential
simplicity is best illustrated in the case of a free non relativistic
particle, discussed here, and can be appreciated by most students taking
an introductory course in Quantum Mechanics. Indeed, the emphasis is on
the role played by the {\it entire family of classical trajectories} in
terms of which the path integral is computed exactly using a functional
representation of the Dirac delta--distribution. We argue that the new
algorithm leads to a deeper insight into the connection between classical
and quantum systems, especially those encountered in high energy physics.
\end{abstract}


\submitted

\maketitle

\section{Introduction}
\label{I}

It is often argued that a formulation of quantum mechanics in
terms of path integrals
is too advanced to lie within the scope of
most undergraduate courses. On the other hand, a great deal of
physics is done, nowadays, using Feynman's path integral method,
with applications ranging from gauge theories  in high
energy physics to solid state
physics, and statistical mechanics. As a result, the amount of
pedagogical literature dedicated to a discussion of the path integral
approach to quantum mechanics is increasing steadily
\cite{one,two,three,four,five}.\\
Quantum mechanics is probably the most challenging paradigm of physics
that a student will ever come across, and the
pedagogical effectiveness of Feynman's method, beautifully expounded
by the author himself in his original work  \cite{six}, lies in its
appeal to intuition  while addressing the most fundamental principles
of the theory. Indeed,
it seems to us that {\it one of the major achievements of Feynman's
formulation of quantum mechanics was to restore the particle trajectory
concept at the quantum level.}
Our purpose, then, is to suggests a specific algorithm that emphasizes
the privileged role that {\it classical} trajectories play in
constructing the ``sum over histories'' envisaged by Feynman in order
to describe the {\it quantum} evolution of a particle.\\
This paper is an outgrowth of an earlier investigation in
string theory where the computational method described
here was first applied
\cite{nine,ten,eleven}.  However, in order to illustrate the method in
the simplest context, we apply it to the familiar case of a free, non
relativistic particle. We show that the path integral can be computed
exactly by summing over the whole family of trajectories which are
solutions of the classical equations of motion. This is in contrast
to the conventional approach to the path integral in which Newtonian
dynamics singles out a unique classical trajectory which is then
perturbed by quantum fluctuations. The main
ideas and techniques are discussed in the case of a free particle
with the bare
minimum of formal apparatus which should be within grasp of most
undergraduate physics majors taking a first course in quantum mechanics.\\
Hopefully,
the applicability of our approach to higher dimensional objects will
justify the claim, made in the abstract, that our method applies to
``simple quantum mechanical systems'' besides the point particle case
considered in this paper. \\
In order to place our work in the right perspective, we begin
Section \ref{II} with a brief overview of the discretization
procedure that forms the basis of the ``sum over histories''
approach to Quantum Mechanics. This will give us the opportunity
to single out the precise point of departure of our own approach
from the conventional one based on the ``Lagrangian path integral''.\\
In Section \ref{III} we construct the propagation kernel and make
contact with the diffusion equation of a free, non relativistic
particle.\\
In \ref{appA} we outline a possible extension
of our definition of path integral
to the case of a particle in an external potential.\\
Finally, \ref{appB} is a compendium of three basic
formulae which are especially relevant to the discussion in the text.

\section{Some Remarks about Path Integration}
\label{II}
In this section we review briefly the definition of phase
space path integral. The main purpose here is twofold: first,
we wish to present a self--contained exposition for those readers
who may not be acquainted with the idea of ``sum over histories''
proposed by Feynman\cite{six,seven}; second, we wish to isolate
those fundamental properties of Quantum Mechanics, encoded in the
path integral approach, which are relevant to the subsequent
discussion of our own computational method . \\
One such fundamental property is the distinction between the
square of the absolute value of the wave function,
or {\it probability density}, and the wave function itself,
or {\it probability amplitude.} With that distinction in mind,
the peculiarity of the quantum mechanical world is that,
in contrast to the classical rules of conditional probability,
one computes probability amplitudes for the paths, and then sum
the amplitudes; when amplitudes are superimposed, and then squared,
an interference pattern is induced in the probability density.
The famed ``particle--wave duality'' of the quantum world stems
essentially from that new computational rule. According to Dirac
and Feynman \cite{twelve}, the basic difference
between classical and quantum mechanics is
that the former selects, through the stationary action
principle,  a single trajectory connecting the initial
and final particle position, while the latter assumes that a
particle ``moves simultaneously along all  possible trajectories''
connecting the initial and final end points. Of course, not all
trajectories are equal, even if they correspond to wave functions
having the same absolute value: the action
$S\left[\,\y(t)\,\right]$ of a particular
path connecting the two fixed end points determines the {\it phase} of the
propagation amplitude associated with that path. The phase factor
corresponding to any individual path can be written in the form
$
 \dis{
 	\exp
 	\left\{
    	\, i \,
        S \left[ \, \y(t) \, , \, \p(t) \, \right]
        /
        \hbar
    \right\}
     }
$,
and a very effective way to
describe the quantum mechanical behavior of a particle is through
the {\it sum over histories} proposed by Feynman.
Symbolically, the amplitude to evolve from $\x_0$ to $\x$ is
\be
    A \left(\, \x_0\, ,  \x\, \right) = \sum_{
                       \parbox{1.2cm}{
                                    \centerline{\tiny paths from
                                     $\x_0$ to $\x$}
                            }
    				  }
                    \Phi \left [\, \y(t) \, , \p (t)\, \right]
    \quad ,
\ee
where
\be
    \Phi \left[ \, \y(t) \, , \p(t) \, \right]
    =
    \K \cdot \exp
    \left\{ \frac{i}{\hbar} S \left[\, \y(t) \, , \p (t)\, \right] \right\}
    \quad .
\ee

A sum
of this type, defined over a space of functions, i.e., $\y(t)$,
and $\p(t)$, is  a {\it functional integral,}
and the basic problem of the Feynman
formulation of quantum mechanics is to determine what that formal sum
means.\\
In order to construct the sum over all paths in phase space,
one may follow the evolution of the system in a total time interval
$[\, t\, ,t'\,]$ by dividing such interval in $N+1$ subintervals, with end 
points
labeled by $t _{1}$, $t _{2}$, \dots , $t _{i}$ , \dots , $t _{N+1}$ so that
\begin{eqnarray}
    t & = & t _{0}
	\nonumber \\
    t' & = & t _{N+1}
    \quad .
\nonumber
\end{eqnarray}
For convenience, one may also choose each of the above intervals to be of
equal length $\epsilon$,
\be
    t _{i+1} - t _{i} = \epsilon
    \quad ,
\ee
in such a way that
\be
    t _{k} = t + k \epsilon \quad , \qquad k = 0 , \dots , N+1 \quad .
\ee
Next, one has to determine an equation for the partial amplitude 
\be
    A \left(\, q _{i+1}\, , t _{i+1}\, ; q _{i}\,  , t _{i}\,\right)
    \equiv
    \left \langle \, q _{i+1}\, , t _{i+1}\, \vert\, q _{i}\,  , t _{i}\, 
    \right \rangle
\ee
that the system evolves from a position with coordinate
$q _{i}$ at time $t _{i}$ to a position with coordinate
$q _{i+1}$ at time $t _{i+1}$.
Using the bracket notation, the amplitude for the first
subinterval takes the form
\begin{eqnarray}
    \left \langle\,
		\tilde{q}\, , \epsilon | \bar{q}\, , 0\,
    \right \rangle
    & = &
    \left \langle\,
		\tilde{q} | \exp \left( - i\, H\, \epsilon \right) | \bar{q}
    \, \right \rangle
    \nonumber \\
    & = &
    \delta \left(\, \tilde{q} - \bar{q} \, \right)
    -
    i \epsilon
    \left \langle\,
        \tilde{q} | H | \bar{q}
    \right \rangle
    + {\mathcal{O}} \left( \epsilon ^{2} \right)
    \nonumber \\
    & = &
    \int \frac{dp}{2 \pi}
        e ^{i p \left( \tilde{q} - \bar{q} \right)}
    -
    \int \frac{dp}{2 \pi}
        e ^{i p \left( \tilde{q} - \bar{q} \right)}
        i \epsilon
        H \left( p , \frac{\tilde{q} + \bar{q}}{2} \right)
    + {\mathcal{O}} \left( \epsilon ^{2} \right)
    \nonumber \\
    & = &
    \int \frac{dp}{2 \pi}
        e ^{i p \left( \tilde{q} - \bar{q} \right)}
        \left [
            1
            -
            i \epsilon
            H \left( p , \frac{\tilde{q} + \bar{q}}{2} \right)
        \right ]
    + {\mathcal{O}} \left( \epsilon ^{2} \right)
    \nonumber \\
    & = &
    \int \frac{dp}{2 \pi}
        \exp
            \left\{
  	          	i
                \left [
                	p \left( \tilde{q} - \bar{q} \right)
    	        	-
        	    	\epsilon
            		H \left( p , \frac{\tilde{q} + \bar{q}}{2} \right)
                \right ]
            \right\}
    \quad .
\end{eqnarray}
Generalizing the above expression to a generic subinterval, we find
\begin{equation}
    \fl
    \left \langle
		q _{i+1} , t _{i+1} | q _{i} , t _{i}
    \right \rangle
    =
    \int \frac{dp _{i}}{2 \pi}
        \exp
            \left\{
  	          	i
                \left [
                	p _{i}
                    \left( q _{i+1} - q _{i} \right)
    	        	-
            		H \left(
                    	p _{i} ,
                        \frac{q _{i+1} + q _{i}}{2}
                      \right)
        	    	\left( t _{i+1} - t _{i}\right)
                \right ]
            \right\}
    \quad .
\label{introonestep}
\end{equation}
Approximating the path from $q$ at time $t$ to $q'$ at time
$t'$ with a finer subdivision of the total time interval,
and in view of the fact that there are no predetermined
conditions on the coordinate variable at intermediate times,
the total amplitude can be written as the product
of the amplitudes associated with each subinterval, integrated
over all possible intermediate positions:
\be
    A \left(\, q'\, , t'\, ; q\, , t \,\right)
    =
    \lim _{N \to + \infty}
    \int d q _{1}
    \int d q _{2}
    \dots
    \int d q _{N}
        \prod _{i} ^{1,N+1}
        \left \langle
            q _{i} , t _{i}
            ;
            q _{i - 1} , t _{i - 1}
        \right \rangle
\quad .
\ee
Combining the two equations above, we are led to the
final step of the discretization procedure
\begin{eqnarray}
    \fl
    A \left(\, q'\, , t'\, ; q\, , t \, \right)
    & = &
    \lim _{N \to + \infty}
    \int d q _{1}
    \int d q _{2}
    \int \dots
    \int d q _{N}
    \cdot
    \nonumber \\
    \fl
    & & \qquad \cdot
    \int \frac{d p _{1}}{2 \pi}
    \int \frac{d p _{2}}{2 \pi}
    \int \dots
    \int \frac{d p _{N+1}}{2 \pi}
    \cdot
    \nonumber \\
    \fl
    & & \qquad \qquad \cdot
    e ^{i \left[\,
              p _{1} \left( \, q _{1} - q _{0} \,\right)
              -
              \left( \, t _{1} - t _{0}\, \right)\,
              H\,  \left(\,  p _{1} , \frac{q _{1} + q _{0}}{2} \,\right)
          \right]
       }
    \cdot
    \nonumber \\
    \fl
    & & \qquad \qquad \qquad \cdot
    e ^{i \left[\,
              p _{2} \left(\, q _{2} - q _{1} \,\right)
              -
              \left(\, t _{2} - t _{1}\, \right)
              H \left(\,  p _{2} , \frac{q _{2} + q _{1}}{2} \,\right)\,
          \right]
       }
    \cdot \dots \cdot
    \nonumber \\
    \fl
    & & \qquad \qquad \qquad \qquad \cdot
    e ^{i \left [\,
              p _{N+1} \left(\, q _{N+1} - q _{N}\, \right)
              -
              \left(\, t _{N+1} - t _{N}\, \right)
              H \left(\, p _{N+1} , \frac{q _{N+1} + q _{N}}{2}\, \right)\,
          \right]
       }
    \nonumber \\
    \fl
    & = &
    \lim _{N \to + \infty}
    \left( \prod _{i} ^{1,N} \int d q _{i} \right)
    \left( \prod _{j} ^{1,N+1} \int d p _{j} \right)
    \cdot
    \nonumber \\
    \fl
    & & \qquad \cdot
    e ^{i \left \{
              \sum _{k} ^{1,N+1}
              \left[
                  p _{k} \left( q _{k} - q _{k-1} \right)
                  -
                  \left( t _{k} - t _{k-1} \right)
                  H \left( p _{k} , \frac{q _{k} + q _{k-1}}{2} \right)
              \right ]
          \right\}
       }
    \nonumber \\
    \fl
    & = &
    \lim _{N \to + \infty}
    \left( \prod _{i} ^{1,N} \int d q _{i} \right)
    \left( \prod _{j} ^{1,N+1} \int d p _{j} \right)
    \cdot
    \nonumber \\
    \fl
    & & \qquad \cdot
    e ^{i \left \{
              \sum _{k} ^{1,N+1}
              \left[
                  p _{k} \frac{q _{k} - q _{k-1}}{t _{k} - t _{k-1}}
                  -
                  H \left( p _{k} , \frac{q _{k} + q _{k-1}}{2} \right)
              \right ]
              \left( t _{k} - t _{k-1} \right)
          \right\}
       }
    \quad .
    \nonumber
\end{eqnarray}
The expression in the last equality represents
the discrete version of the path integral in phase space
\begin{equation}
    \int \left[ {\mathcal{D}} q \right]
    \int \left[ {\mathcal{D}} p \right]
        \exp \left\{
                i
                \left [
                    p \, \dot{q}
                    -
                    H \left( p , q\right)
                \right ] dt
             \right\}
    \quad .
\label{introdefpathint}
\end{equation}
which is the starting point for computing
the sum over histories for the majority of
the integrable systems encountered in the
literature. Equation (\ref{introdefpathint})
is also the starting point of our own approach.
However, in terms of the discretization procedure
outlined above, a further simple elaboration of
the phase space path integral leads to a computational
method which seems mathematically more efficient and
physically more enlightening, at least for some complex
systems encountered in high energy physics.\\
Consider the first term in the expression (\ref{introdefpathint}), namely
\be
    \int p\, \dot{q} \, dt = \int p\, dq 
    \quad .
\ee
This term corresponds to the discrete sum
\be
    \sum _{k} ^{1,N+1}
        p _{k} \left( q _{k} - q _{k-1} \right)
\ee
which we can rewrite as follows:
\begin{eqnarray}
    & &
    \sum _{k} ^{1,N+1}
        p _{k} \left(\, q _{k} - q _{k-1}\, \right)
    =
    \nonumber \\
    & & =
    \sum _{k} ^{1,N+1}
        p _{k} q _{k}
    -
    \sum _{k} ^{1,N+1}
        p _{k} q _{k-1}
    \nonumber \\
    & & =
    \sum _{k} ^{1,N+1}
        p _{k} q _{k}
    -
    \sum _{k} ^{1,N+1}
        p _{k} q _{k-1}
    -
    \sum _{k} ^{1,N+1}
        p _{k-1} q _{k-1}
    +
    \sum _{k} ^{1,N+1}
        p _{k-1} q _{k-1}
    \nonumber \\
    & & =
    \sum _{k} ^{1,N+1}
        \left(
            p _{k} q _{k}
	    -
            p _{k-1} q _{k-1}
        \right)
    -
    \sum _{k} ^{1,N+1}
        \left(
            p _{k} - p _{k-1}
        \right)
        q _{k-1}
    \nonumber \\
    & & =
    p _{N+1} q _{N+1}
    -
    p _{0} q _{0}
    -
    \sum _{k} ^{1,N}
        \left(
            p _{k} - p _{k-1}
        \right)
        q _{k-1}
    \quad .
\end{eqnarray}
The first two terms in the above expression represent a {\it boundary
contribution} to the path integral, whereas the
remaining sum corresponds to the following integral
\be
    \sum _{i} ^{1,N}
         \left( p _{k} - p _{k-1} \right) q _{k-1}
    \Longrightarrow
    \int q \, dp
    =
    \int q \, \dot{p} \, dt
    \quad .
\ee
Next, in what follows we make use of the following equality
\be
    \int d q _{k-1}
        \exp \left\{ i \left( p _{k} - p _{k-1} \right) q _{k-1} \right\}
=    \delta \left( p _{k} - p _{k-1} \right)
\ee
which gives the well known representation of the Dirac delta function as
Fourier transform of the imaginary exponential.
Thus, the net result of the above rearrangement
of terms is encoded in the following correspondence
\be
    \fl
    \int \left [ {\mathcal{D}} q \right]
        \exp \left\{ i \int dt\, q\, \dot{p} \right\}
    =
    \delta \left[ \,\dot{p} \,\right ]
    \quad \Longrightarrow \quad
    \lim _{N \to \infty}
    	{\mathcal{N}} \left( t _{j} \right)
    	\prod _{i} ^{1,N}
            \delta \left(
          		 	  \frac{p _{k} - p _{k-1}}{t _{k} - t _{k-1}}
            	   \right)
    \quad .
\ee
which we take as a definition of the
``functional Dirac delta distribution''.\\
In the next section, we apply this mathematical
rearrangement of terms in the phase space path
integral to compute the propagation kernel of a
non relativistic particle. As we shall see, the
physical payoff is a novel interpretation of the
sum over histories in the sense that it underscores
the special role played by the entire family of
trajectories which are solutions of the classical
equations of motion.

\section{Non-Relativistic Particle}
\label{III}

\subsection{Path Integral and Propagation Kernel}
\label{IIIA}

The classical dynamics of a
non-relativistic point--particle with mass $m$
is encoded into the Lagrange function, or in its corresponding
Hamiltonian
\bee
&& L\left(\, \y(t) \, , \, \vy(t) \, ; \, t\, \right)
={1\over 2}\,m\, \vy{}^2 \\
&& H\left(\, \y(t) \, , \, \p(t) \, ; \, t\, \right)={1\over 2 m}\,\p{}^2
\quad .
\eee
>From here follow the classical equations of motion:
\bee
{d\p\over dt}=0 \quad \Rightarrow \quad \p={\rm cost.}\equiv \q \label{pcl}
 \\
{dH\over dt}=0 \quad \Rightarrow \quad H={\rm cost.}\equiv E
 \\
m{d\y\over dt}=\q \quad \Rightarrow \quad \y(t)={1\over m}\,\q \, t+\x_0
\quad ,
\eee
where we have taken into account the boundary conditions
\be
\y(0)=\x_0 \quad,\qquad \y(T)=\x \quad \Rightarrow \quad \q=m\,{\x-\x_0\over
T}
\ee
so that
\be
\y(t)={\x-\x_0\over T}\, t+\x_0
\quad .
\ee
Furthermore, for later reference, we recall that the classical action
\bee
\scl(\, \x\, , \, \x_0 \, ; \, T \,)&=&{1\over 2}m\int_0^T dt\,
\left(\, {\x-\x_0\over T}\, \right)^2\nonumber\\
&=&{m\over 2T}\vert\,\x-\x_0\, \vert^2
\nonumber
\eee
is a solution of the Jacobi equation
\be
{\partial\scl\over\partial T} +
{1 \over 2m} \vec\nabla\scl\cdot\vec\nabla\scl = 0
\quad .
\label{hj}
\ee
Equivalently, one can follow the particle
evolution by determining the
propagation amplitude $K(\, \x -\x_0 \, ; \, T \, )$
that a particle propagates from an initial
position $\x_0$ to a final position $\x$. As we
have shown in the previous section, the amplitude
is given by a {\it formal}
sum over all  phase space paths connecting
$\x_0$ to $\x$ in a
total time $T$, each path carrying a weight
given by the phase factor
$\exp\left(\, i\, S[\, \hbox{path} \, ]/\hbar\, \right)$.
We have also reviewed the standard method that
gives meaning to the sum over histories: it goes
through a  discretization
procedure which consists in subdividing the
total time lapse $T$ into
a number of infinitesimal intervals, thereby
approximating the smooth phase
space trajectory followed by a classical
particle with a succession of
``jagged'' paths. It seems worth emphasizing
at this point, that in the above implementation of
the path integral, the classical trajectory followed
by a particle is uniquely specified by Newtonian mechanics.
Furthermore, we hasten to say that the sum over histories,
so defined, is a sum over trajectories which are
nowhere differentiable. As a matter of fact, a post
modern interpretation of Feynman's discussion is that
the quantum mechanical path of a particle is inherently
{\it fractal}\footnote{This interpretation, clearly
implied in the book by Feynman and Hibbs\cite{seven},
was revisited many years ago by Abbott and Wise\cite{eight},
and reviewed by the authors in the larger, but mostly
unexplored context of the microscopic structure of quantum
spacetime\cite{nine,ten}.}. The gist of the argument is
that, when a particle is more and more precisely located
in space, its trajectory becomes more and more erratic as
a consequence of Heisenberg's principle. In other words,
it is the addition of quantum fluctuations
around the classical trajectory that
gives meaning to the idea that ``a particle
moves along all possible paths'' connecting
the initial and final configurations. Be that
as it may, the discretization procedure turns
the functional integral  into an
infinite product of ordinary integrals
\cite{thirteen}. For some
elementary systems, the  integration over the momenta can be carried
out, albeit with some efforts, and the final result
for the amplitude is the ``Lagrangian path integral'':
\be
K(\, \x \, , \, \x_0 \, ; \, T \,)
=\int_{\x_{0}}^\x[{\mathcal{D}}y(t)]\exp{i\over\hbar}\int_0^T
L[\,\y(t) ,\vy(t)\, ; t\,]\, dt \quad .
\label{kuno}
\ee
For more complex systems, such as the
relativistic extended systems encountered
in contemporary high energy physics, the
expression (\ref{kuno}), may not tell the
full story, and we have found it advantageous
to start with a sum over histories in phase space.
Thus, in order to illustrate our computational method,
we start directly from the
canonical phase space path integral for a free, non--relativistic
particle
\be
\fl
 K(\, \x \, , \, \x_0 \, ; \, T \,)=N\int_{\x_0}^{\x}[{\mathcal{D}}y(t)]
 [{\mathcal{D}}p(t)]\exp
{i\over\hbar}\int_0^T dt\,\left[\, \p(t)\cdot{\vy}(t)-
H_0(\p)\, \right]
\quad ,
\ee
where
\be
 H_0(\p)={\p{}^2\over 2m}
\ee
and $N$ is a normalization constant to be fixed later on.\\
The focal point of our approach is the recognition that the path
integral automatically assigns a special role to
the {\it whole family} of trajectories
which are solutions of the classical equation
of motion. The precise meaning of the above statement
is illustrated by implementing the computational steps
following Eq.(\ref{introdefpathint}).
Thus,
we write the first term in the action as
\be
\fl
\int_0^T dt\,\p\cdot{\dot\y}=\int_0^T dt\, {d\over dt}\left[\, \p\cdot\y
\, \right]-
\int_0^T dt\, \y\cdot {\dot\p}
 =\int_{x_0}^x d\left(\, \p\cdot\y\, \right)-\int_0^T dt\, \y\cdot {\dot\p}
\quad .
\label{int}
\ee
Accordingly, the path integral reads
\bee
\fl
&&   K(\, \x \, , \, \x_0 \, ; \, T \,)=\nonumber\\
\fl
&&   \quad = N\int_{\x_0}^{\x}[{\mathcal{D}}y(t)][{\mathcal{D}}p(t)]
\exp\left\{
{i\over\hbar}\left[\int_{\x_0}^\x d\left(\p\cdot\y\right)
-\int_0^T dt\, \y(t)\cdot {\dot\p}(t)
-\int_0^T dt\, H_0(\p)\right]\right\}
\nonumber
\eee
and we note that, while the first term in the above expression is the
integral
of a total differential, and therefore independent of the path
connecting the two points $\left(\, \x_0\, , \x\,\right)$, the second term
depends \textit{linearly} on the spatial trajectory $\y(t)$.
A closer look at this term shows that it is a (functional) Dirac--delta
distribution represented as a ``Fourier integral'' over the functions
$\y(t)$
\be
\int_{\x_0}^{\x}[{\mathcal{D}}y(t)]
\exp\left\{- {i\over\hbar}\int_0^T dt\, \y(t) \cdot {\dot\p}(t)\right\}
=\delta\left[\, {d\p\over dt}\,\right]
\quad .
\label{dirac}
\ee
The Dirac--delta in equation (\ref{dirac}) is non-vanishing only when its
argument is zero, i.e., when the momentum $\p$ solves the
classical equation of motion
\be
{d\p\over dt}=0 \quad \Rightarrow \quad \p(t)={\rm const.}\equiv \q
\quad .
\label{eqcl}
\ee
This is our  first, and central result which applies equally well,
``mutatis mutandis'', to relativistic point--particles and relativistic
extended objects: {\it once the spatial
trajectories} $\y(t)$ {\it are integrated out,
the resulting path integral is
non-vanishing only when the surviving integration variable, namely
the three--momentum vector along the trajectory,  is constrained to
satisfy equation}{ }(\ref{eqcl}). This information
is encoded in the following
expression for the propagation amplitude
\bee
\fl
&&
K(\, \x \, , \, \x_0 \, ; \, T \,)=
\nonumber \\
\fl
&& \quad =
N
\int [{\mathcal{D}}p(t)] \,
\delta\!\left[\frac{d\p}{dt}\right]
\exp\left[ {i\over\hbar}\int_{\x_0}^\x \! \! \! d\left(\,
\p\cdot\y\, \right)\right]\,
\exp\left\{-{i\over\hbar}\int_0^T \! \! \! dt H_0(\p)\,\right\}
\quad ,
\label{pint}
\eee
where only the restricted {\it family of constant (i.e., time
independent) $\p$--trajectories} contributes to the path integral.
Note that the value of the three--momentum along a classical
trajectory is a fixed, even though arbitrary, {\it number}.
Thus, on the mathematical side,
the pay--off of our procedure is that we 
have traded the original path integral, i.e., a functional integral, with
a single ordinary integral over the constant, but
numerically arbitrary, components of the three--momentum.\\
On the physical side, our method underscores a conceptual
difference between the mechanism of propagation envisaged
here and the conventional one. By this we mean that in
the conventional interpretation of the path integral,
the sum over histories is obtained by summing over all
possible quantum fluctuations around a {\it single}
classical trajectory which is uniquely defined because
{\it both} extremal position and momentum are preassigned.
In contrast, in  our interpretation of sum over histories, only the
extremal {\it coordinates} of the particle are precisely specified by the 
boundary conditions, whereas the corresponding value
of the momentum at the end points
is constant but {\it arbitrary}, in consistency with
the uncertainty principle. Since the momentum is a vector, it follows
that all paths in configuration space contribute to
the evolution of the wave
function. In the above sense, it seems to us that
the complementarity between particle and wave behavior
and the role of the uncertainty principle are
especially evident in our approach.\\
To summarize our discussion so far, ``the sum over
histories'' of a free particle between two fixed end points, can be
translated into the integration over all possible values of the linear
momentum. Put briefly,
\bee
 \int [{\mathcal{D}}p(t)] \,\delta\left[d\p/dt\right]\left(\dots\right)
&=&{\rm sum\  over\  the\  classical\  momenta}
\nonumber\\
  &=&{\rm sum\ over\ constant\ momentum\ trajectories}
\nonumber\\
  &=&\int d^3 q \left(\dots\right) \equiv {\rm ordinary\ momentum\ integral}
\quad .
\nonumber
\eee
Next, if the momentum is constant, then the Hamiltonian is constant
as well, and we can write
\be
\fl
\int d^3 q \,\exp\left[ {i\over\hbar}\q\cdot\int_{\x_0}^\x d\y
-{i\q {}^2\over 2m\hbar}T \right]=
\left[{\rm det}\left({i T\over 2\hbar m}\delta_{ij}\right)
\right]^{-1/2}
\exp\left[{i m\over 2\hbar T}\vert\, \x-\x_0\, \vert^2\right]
\quad ,
\ee
where we have used the identity
\be
\dis{\int_{\x_0}^\x d\vec{\bf y}=\x-\x_0}
\ee
and the formula for the multidimensional Gaussian
integral listed in Appendix
\ref{appB}. As a result, we obtain
\be
K(\, \x \, , \, \x_0 \, ; \, T \,)
=N\,\left({2\hbar m\over    i\pi T }\right)^{3/2}
\exp\left[{i m\over 2\hbar T}\vert\, \x-\x_0\, \vert^2\right]
\quad .
\ee
Finally, the normalization constant $N$ can be fixed by the condition
that, in the limit of a vanishing lapse of time,
the particle is bound to be in its initial position. In other words,
\be
\lim_{T\rightarrow 0} K(\, \x - \x_0\, ; \, T \, )
=
\delta^3\left(\, \x-\x_0\, \right)
\quad
\Rightarrow \quad N=\left(\, {1\over 2\pi\hbar^2}\, \right)^{3/2}
\quad .
\label{bc}
\ee
Hence,
\bee
K(\, \x -\x_0 \, ; \, T \,)
&=&
\left({m\over 2i\pi\hbar T }\right)^{3/2}
\exp\left[{im\over 2\hbar T}\vert\, \x-\x_0\, \vert^2\right]\nonumber\\
&=&\left({m\over 2i\pi\hbar T }\right)^{3/2}
\exp\left[\, i\, \scl\left(\, \x - \x_0 \, ; \, T \,\right)/\hbar\, \right]
\quad ,
\label{free}
\eee
where the phase factor is just the classical action measured in $\hbar$
units
\be
\scl\left(\, \x - \x_0 \, ; \, T \,\right) = {m\over 2} \int_0^T dt\,
\left(\, {\x-\x_0\over T}\, \right)^2=
{m\over 2T}\vert\, \x-\x_0\, \vert^2
\quad .
\ee
The above expression for the propagation kernel is
a well known result which tests the consistency of our approach:
as Feynman and Hibbs demonstrated, {\it for any lagrangian quadratic in
the position and velocity variables, the corresponding
propagation amplitude is given by a pre--factor, which is a function
of the evolution parameter lapse, multiplied by the exponential of ``i''
times the classical action in $\hbar$ units.}

\subsection{Diffusion Equation and Propagation Kernel}
\label{IIB}

 In the previous subsection, we have deduced the form
 of the propagation kernel (\ref{free}), by
evaluating the sum over histories as a Gaussian integral.
With this result in hands, one can make contact with the
more familiar formulation of quantum mechanics by showing that the
propagation kernel $K$ satisfies a diffusion equation of the
Schr\"odinger type. To see this, first we calculate
\be
\fl
{\partial\over\partial T}K(\, \x - \x_0 \, ; \, T \,)=N\,
\int d^3 q \, \left(-i{\q \cdot\q \over 2\hbar m}\right)
\exp\left[ {i\over\hbar} \q \cdot \int_{\x_0}^\x d\y \right]\,
\exp\left\{-{iT\over2\hbar m} \q \cdot \q \right\}
\quad ,
\label{eq1}
\ee
and
\be
\fl
\Delta_x K(\, \x -\x_0 \, ; \, T\, )=N\,
\int d^3 q \, \left(-{\q \cdot \q \over \hbar^2}\right)
\exp\left[ {i\over\hbar}\q \cdot\int_{\x_0}^\x d\y\right]\,
 \exp\left\{-{iT\over2\hbar m}\q \cdot \q\right\}
\quad .
\label{eq2}
\ee
Comparing the two equations above, we conclude that
\be
{\hbar^2\over 2m}\Delta_x K(\, \x - \x_0 \, ; \, T \,)=
i\hbar{\partial\over\partial T}K(\, \x - \x_0 ; T \,)
\quad .
\label{diff}
\ee
Finally, recalling the relation between the amplitude $K$ and the wave
function, namely
\be
\psi(\,\x \, , \, t)=
\int d^3 x_0 \, K(\, \x -\x_0 \, ; \, t \,)\,\psi(\x_0 \,
,
\,0)
\quad ,
\ee
we also conclude that $\psi(\,\x\, , t\,)$ must satisfy the time--dependent
Schr\"odinger equation.\\
 At this point, it
seems pedagogically instructive to reverse the procedure, and show that the
propagation kernel can be determined by solving the diffusion equation
(\ref{diff}) which we now assume as given.
To this end, we make the following ansatz,
\be
K(\, \x -\x_0 \, ; \, T \,)
= F(T)\,
\exp\left( \, i\, M(\, \x - \x_0 \, ; \, T \,)/\hbar\, \right)
\quad ,
\label{ansatz1}
\ee
in terms of two trial functions $F(T)$ and $M(\, \x -\x_0 \, ; \, T\, )$.
The overall normalization constant may be determined by the same
boundary condition (\ref{bc}).\\
In order to determine the form of the trial functions, we demand that the
tentative expression (\ref{ansatz1}) satisfies equation (\ref{diff}).
Thus, one
finds
\be
{F(T)\over 2m}\, \Delta_x M (\x -\x_0 \, ; \, T)=-{ d F(T)\over d T}
\label{cont}
\ee
and
\be
{1\over 2m}\left(\vec\nabla M\right)\cdot
\left(\vec\nabla M\right)=-{\partial M\over \partial T}
\quad .
\label{jacob}
\ee
Note that equation (\ref{jacob}) is just the classical
Jacobi equation (\ref{hj}). Thus, without further
calculations we can identify the phase of the
kernel with the classical action:
\be
M(\, \x -\x_0 \, ; \, T \,)=\scl\left(\, \x -\x_0 \, ; \, T \,\right)
=
{m\over 2T}\vert\,\x-\x_0\vert^2
\quad .
\ee
Next, in order to determine the dependence of the kernel pre--factor on
$T$, we make use of equation (\ref{cont}). Therefore, we first apply the
Laplacian operator to $\scl(\, \x -\x_0 \, ; \, T \,)$:
\bee
&&{\partial\over\partial x^i}\scl(\, \x -\x_0 \, ; \, T \, )= {m\over T}
\left(\x-\x_0\right)_i  \\
&&{\partial^2 \over \partial x^k \partial x^i}
\scl(\, \x -\x_0 \, ; \, T \,)
= {m\over T}\delta_{k\,i} \\
&&\delta^{k\,i}{\partial^2\over   \partial x^k  \partial x^i}
\scl(\, \x -\x_0 \, ; \, T \,)= {m\over T}\delta^k{}_k={3m\over T}
\quad .
\eee
Then, we substitute the result into equation (\ref{cont}),
which now takes the form
\be
{dF\over dT}=-{3\over 2T}F \quad\Rightarrow \quad
{dF\over F}=-{3\over2}{dT\over T}
\quad .
\ee
Finally, integrating the last equation, we obtain
\be
F(T)={\rm const.}\times T\,{}^{-3/2}
\quad .
\ee
Substituting all of the above in the original ansatz (\ref{ansatz1}),
leads to the following expression for the propagation kernel,
\be
K(\, \x -\x_0 \, ; \, T \,)=
{N\over T^{3/2}}
\exp\left(\, i\, \scl(\, \x -\x_0 \, ; \, T \,)/\hbar\, \right)
\quad .
\ee
Except for the normalization constant which is fixed by the initial
condition, this is the same expression obtained from the path integral.

\section{Concluding Remarks}

As argued  in Ref. \cite{three}, the conceptual foundations of
quantum mechanics are still the subject of debate, and thus ``it is
important to introduce undergraduate physics majors to non-standard
developments of quantum mechanics'' in the hope that exposure to such
developments, early in the curriculum, might motivate some students
to focus on the physical principles underlying the theory.
In this paper we have taken up that challenging task by
discussing a novel approach to the computation of the
sum over histories in the path integral.  \\
 On the mathematical side, we have suggested a phase space formulation
of the Feynman sum over paths in which the spatial
trajectory $\y(t)$ plays the role of a Lagrange
multiplier enforcing the classical equation of
motion (\ref{pcl}) at the quantum level. Imposing such
a constraint, with the aid of a Dirac delta--distribution, reduces the
functional integral to an ordinary integral over the arbitrary values of
the initial momentum. In the simple case of a free,
non relativistic particle, the use of the phase space
path integral, or the Lagrangian formulation of it, is
largely a matter of choice, in the sense that one arrives
at the same expression of the propagation kernel. In the
case of relativistic extended systems, the use of the
phase space path integral along the same lines discussed
here, seems definitely advantageous, if not mandatory.
On the physical side, however, our formulation of the
path integral reflects a new mechanism of propagation
which, in our view, sheds some new light on the connection
between classical and quantum systems. The element of
novelty can be summarized thus: it is well known that
classical mechanics, which is a deterministic theory,
demands that one definite value of the initial position
and momentum be assigned a priori, and thereafter selects a
{\it single classical trajectory} that satisfies
the equation of motion. On the other hand, in quantum
mechanics one cannot  fully specify the value of the
momentum of a precisely localized particle. The price to pay
for specifying the particle position at $\x_0$
when $t=0$, is that  $\p$ is completely
arbitrary, and therefore one has to sum over  all
possible values of the three--momentum. This brings
into the game the whole family of trajectories which
satisfy the classical equations of motion. With all
such channels of propagation open, a quantum particle
turns to a ``wave--like'' propagation mode, exploring
all of them by
 ``evolving simultaneously'' along all classical trajectories
corresponding to all possible values and orientations
of the momentum, up to the final point
 $\x$ where the particle is detected. Thus, as stated
 in most textbooks of Quantum
 Mechanics, but never fully clarified, a quantum
 particle starts as a pointlike
 object precisely located at a point, propagates as
 a wave, only to reappear as a dot on a screen, at
 the point of detection. We have exploited this
 {\it dual, } wavelike, description of the family
 of classical trajectories by connecting the quantum
 evolution of a particle with the Jacobi formulation
 of Classical Mechanics, i.e., by deriving the diffusion
 equation from the propagation kernel, and vice versa.
 The main purpose of the whole exercise was to test the
 consistency of our approach, and its ability to reproduce
 some standard results obtained by different methods.
 Equally important, however, the objective was to clarify,
 with the familiar example of a free particle, the connection between the
classical
theory and the central idea upon which the path integral
formulation is
based,
namely, that ``trajectories'' still play a
role in quantum mechanics and that
all paths, with their intrinsic randomness,
contribute to the evolution of the
wave function which is governed by the propagation
kernel. Having said that, we may add that our
mathematical analysis and its physical interpretation
extend well beyond the case of a free, non relativistic particle.
Indeed, the very possibility of formulating a quantum mechanics
of closed strings and other extended objects\cite{eleven},
points to the effectiveness of our approach to the
path integral for studying new fundamental physics.
By an interesting feedback process, we have found that
many new formal and physical properties of our approach
to strings can be applied equally well to the case of a
point particle which is now regarded as the object of
lowest dimensionality in a hierarchy of geometric objects
to which the same dynamical principles seem to apply.
>From the present introduction,
the interested reader may eventually
progress to a more advanced discussion involving relativistic extended
systems, such as strings, membranes and p--branes, with a change in
notation rather than substance. 

\appendix
\section{Particle in an External Potential: Short Distance Limit}
\label{appA}

The results obtained in the previous sections for a free particle
are, of course, {\it exact}. However,
the same approach can be used to derive the {\it short distance
approximation} for the particle's kernel in an arbitrary potential
$V(\x)$. \\
If $|\x-\x_0|$ is much smaller than the range over which $V(\x)$
varies significantly, then we can Taylor expand $V(\x)$ in the
neighborhood of $\x_0$:
\bee
& &
V(\x) \approx  V_0 -\F \cdot (\x -\x_0)+\dots \\
& &
\F = -\left[ \vec{\nabla}  V(\x) \right \rceil _{\x=\x_0}
\quad .
\eee
In this case, the path integral reads
\bee
\fl
&&
K(\, \x - \x_0 \, ; \, T \,)=N\int_{\x_0}^{\x}[{\mathcal{D}}y(t)]
[{\mathcal{D}}p(t)]\times
\nonumber\\
\fl
&&  \quad \times  \exp\left\{\,{i\over\hbar}\left[\,
\int_{\x_0}^\x d\left(\, \p\cdot\y\, \right)
-\int_0^T dt\, \y(t)\cdot{d\p(t) \over dt}
-\int_0^T dt\,\left(\, H -\F \cdot (\y -\y_0)\,\right)\,\right]
\,\right\}
\quad ,
\eee
where $H\equiv H_0+V_0$. Integration over the particle trajectory
gives a ``Dirac--delta'' whose argument yields the classical equation of
motion
\be
{d\p \over dt}=\F
\ee
which leads, in turn, to the expression for the classical momentum
\be
\p(t)=\F t + \q \quad .
\ee
Thus, the integration over the momentum trajectory can be defined as
follows
\be
\fl
  \int [{\mathcal{D}}p(t)]\,\delta\left[\,{d\p\over dt}-\vec F\,\right]
  \left(\dots\right)
\equiv \int d^3 q \int [{\mathcal{D}}p(t)]\,\delta
\left[\,\p(t) - \F t - \q \,\right]
\left(\dots\right)
\ee
and the resulting path integral takes the form
\bee
\fl
K(\, \x - \x_0 \, ; \, T \,)
&=&
\left(\, {1\over 2\pi\hbar^2}\, \right)^{3/2}
\exp {i \, T\over \hbar}\left(\, V_0 - \F \cdot \x +
{\F\,{}^2 T{}^2\over 6m}\, \right)\times\nonumber\\
\fl
&& \qquad \qquad \times
\int d^3 q \exp\left(-{i\, T\, \q\,{}^2\over 2\hbar m}\,\right)
\exp\left[ -{i\over \hbar}\q\left(\,\x - \x_0 +{T^2\over 2\hbar m}
\F\, \right)\,\right]\nonumber\\
\fl
&=&\left(\, {m\over 2i\pi\hbar T}\, \right)^{3/2}
\exp {i\, T\over \hbar}\left(\, V_0 - \F\cdot \x +
{\F\,{}^2 T{}^2\over 6m}\, \right)
\times
\nonumber \\
\fl
& & \qquad \qquad \times
\exp\left[-{i\, m\, T\over 2\hbar\, T}
\left(\, \x - \x_0 +{T{}^2\over 2\hbar\, m}\F\,\right)^2\, \right]
\quad ,
\eee
where the normalization constant has been chosen in such a way as to
reproduce equation (\ref{free}) once the potential is switched off.

\section{Three Basic Formulae}
\label{appB}

One needs only three basic formulae in order to derive the main results
reported in this paper: one
is the extension of the Gaussian integral to a $D$--component
vector variable, and the other two are representations
for the Dirac delta--function.

\newlength{\mylength}
\addtolength{\mylength}{\textwidth}
\addtolength{\mylength}{-1cm}

\begin{center}
{\large Generalized Gaussian Integral}\\
\fbox{
\parbox{\mylength}{
\bee
\fl
& &
\int d{\bf X}
\exp\left(-{1\over 2}\, X^i\, A_{ij}\, X^j + B^i\, X_i\, \right)
=
\nonumber \\
& & \qquad \qquad \qquad
=
{\rm const.}\times \left(\, {\rm det}{\bf A}\, \right)^{1/2}
\exp\left(\,
          \vec{\bf B}
          \cdot
          {\bf A}^{-1}
          \cdot
          \vec{\bf B}/2
   \, \right)
\eee
$\>\>$ where
\be
    {\bf A} = \left(\, A  _{ij}\, \right)
    \quad {\rm and} \quad
    \vec{\bf B} = \left(\, B _{i} \, \right)
\ee
}
}
\end{center}

\begin{center}
{\large Dirac--delta representations}\\
\fbox{
\parbox{\mylength}{
\bee
\fl
&&\! \! \! \! \! \! \! \hbox{a) Functional Fourier Transform:}\nonumber\\
&&\delta[\vec{\bf X}(\lambda)]=({\rm{}const.})\times
\int [{\mathcal{D}}P(\lambda)]
\exp\left(-i\int d\lambda\, \vec{\bf P} (\lambda) \cdot
\vec{\bf X} (\lambda)\, \right)
\eee
\bee
\fl
&&\! \! \! \! \! \! \! \hbox{b) Gaussian Representation:}\nonumber\\
&&\delta^D(x)\equiv \lim_{\epsilon\rightarrow 0}
\left(\, {1\over\pi\epsilon}\, \right)^{D/2}
\exp\left(-x^2/\epsilon\,\right)
\eee
}
}
\end{center}

\end{document}